\documentclass[prb,twocolumn,showpacs]{revtex4}
\usepackage{graphicx}
\usepackage{amsmath}
\def\Journal#1#2#3#4{{#1} {\bf #2}, #3 (#4)}

\def\PRL{Phys. Rev. Lett.}

\def\PRB{Phys. Rev. B}

\def\Science{Science}

\def\PhysicaA{Physica A}
\newcommand{\n}{\nonumber}
\newcommand{\bn}{\begin{eqnarray}}
\newcommand{\en}{\end{eqnarray}}
\newcommand{\eml}{\end{multline}}
\newcommand{\bml}{\begin{multline}}
\newcommand{\h}{\hspace}

\begin{document}

\title {A Quantum Singlet Pump}
 \author{Kunal K. Das}
 \affiliation{Department of Physics, Fordham University, Bronx, New York 10458, USA}

\date{\today }
\begin{abstract}

We provide provide a detailed study of biasless coherent transport
of singlet electron pairs in one-dimensional (1D) channels induced
by electron-electron interactions that are time-varying in certain
spatially localized regions of the channel. When the time
variation is cyclic, the mechanism is analogous to the adiabatic
quantum pumping of charge and spin previously studied. However,
the presence of interactions that vary only in localized regions
of space requires an intrinsically two-body description which is
irreducible to the 1D single particle scattering matrix elements
that are sufficient to describe quantum pumping of charge and
spin. Here we derive a generalized theory for the pumping of such
interacting pairs starting from first principles.  We show that
the standard description of charge pumping is contained within our
more broadly applicable expressions. We then apply our general
results to a concrete lattice model and obtain an exact analytical
expression for the pumped singlet current. We further demonstrate
that such a model can be implemented with a chain of currently
available quantum dots with certain minor modifications that we
suggest;  we present a detailed numerical feasibility analysis of
the characteristics of such experimentally realizable quantum
dots, showing that the requirements for a measurable pumped
singlet current are within experimental range.

\end{abstract}
 \maketitle

\section{Introduction}

Pumping mechanisms have attracted research attention in recent
years, both experimental \cite{Kouwenhoven, Pothier, Switkes,
Hohberger, Watson, Dicarlo, Buitelaar, Kaestner}  and theoretical
\cite{Brouwer-1, Das-PRL, Kim-PRB, Aleiner, Samuelsson, Altshuler,
Mucciolo, Levinson-turnstile, nano-transport, Hu-Sarma,
Floquet-Buttiker, Floquet-Kim, Aono, Blaauboer,Beenakker}, because
they provide a remarkable alternative means of generating current in
nanostructures.  Such mechanisms use time varying parameters or
potentials to achieve the flow, rather than the application of a
bias.  While the basic idea behind such a pump follows from
classical intuition, in nanostructures quantum mechanics can
introduce novel features that make ``quantum pumping'' a rich and
non-trivial process.  For example, a quantum dot with time varying
couplings to pinched off leads will pump electrons in a process that
is analogous to the classical pumping of water through a lock in a
canal \cite{Kouwenhoven, Pothier};  here Coulomb blockade plays the
role of gravity, limiting the amount of charge that can flow through
the dot each pumping cycle.  However, in an open quantum dot system,
time varying parameters lead to changes in the electron wavefunction
\cite{Thouless,Avron-Math}, leading to modulation of the spatial
probability density that results in a net transfer of particle
density through the region of time variation.  This quantum
mechanical process is subtle and cannot be understood with classical
intuition alone.

The concept of quantum pumping is a very broad one, generically
relevant when a channel has time varying parameters.  As a result,
one can imagine using pumping to transport any quantity associated
with quantum mechanical wavefunctions that can be varied in time
with relative ease.  Charge \cite{Dicarlo, Buitelaar,Brouwer-1,
Aleiner, Levinson-turnstile, nano-transport} and spin
\cite{Watson,Kim-PRB, Mucciolo,Aono} have been primarily
considered for pumping. These quantities can be associated with
single particles. Their pumping is most often computed within an
independent particle approximation, although there have been
studies that have computed their pumping in a strongly correlated
system like a Luttinger liquid
\cite{Luttinger-Chamon,Luttinger-Niu}.

In a recent paper \cite{Das-PRL}, we proposed the quantum pumping of
a two particle quantity - electron singlet pairs.  This is relevant
from an applications standpoint as well as a fundamental physics
point of view.  A pair of electrons in a singlet state is
spin-entangled \cite{Bohm}, and coherent transport of entangled
states is essential for quantum information \cite{Nielsen-Chuang}.
Describing the pumping of a two particle quantity is quite different
from the usual quantum pumping theory of single particle quantities.
As one of the two main topics of this paper, we derive a theory that
is capable of describing singlet pumping and also incorporates
standard quantum pumping cases.  The inherent presence of two body
interaction in pumping singlet pairs makes the theoretical
description significantly more involved than the commonly used
models based upon single particle scattering matrix elements
\cite{Brouwer-1}. Thus a primary goal of the paper is to provide a
detailed analysis of the necessary theory for a singlet pump.

The second topic of this paper is a study of the experimental
feasibility of a quantum singlet pump based upon a detailed
numerical analysis of an experimentally realizable system
involving quantum dots.  We find that, with laboratory systems
already available, the implementation of a singlet pump could be
possible.

The paper has roughly three parts.  The first part involves the
derivation of our approach to quantum pumping and comprises of Sec.
II, III, and IV.  Section II defines a singlet current and
interacting two particle states.  Section III develops the theory of
a singlet pump starting from an adiabatic perturbation expansion.
These sections rely on Green's functions theorems appearing in
Appendices A and B. Section IV demonstrates that our results recover
the established theory when one considers quantum charge pumping of
single electrons. The second part of the paper appears in Section V
which describes a physical model based upon a two particle analog of
a turnstile, involving a chain of quantum dots. An analytic
expression for the singlet current in that model is derived using
the Green's function theorems in the appendices once again. The
third and last part comprising of Sec. VI, VII and VIII is devoted
to a feasibility analysis of a proposed experimental implementation
of a singlet pump based upon the physical model presented in Sec. V.
Section VI suggests use of a specific design of quantum dot similar
to an experimentally realized dot, and the spatial potential energy
profile of an electron in the dot is numerically computed. Section
VII evaluates the energy of two electrons in such a dot, and finally
Sec. VIII demonstrates that such a dot has the features necessary
for generating a pure singlet current while suppressing current of
single particles and triplet pairs.

\section{Singlet Current}

\subsection{Definition of the current}

The quantum mechanical current density is generally defined from the
continuity equation by considering the time variation of the single
particle density. In order to discuss quantum pumping of singlets,
we need to have an appropriate definition of a singlet current.  We
thus begin with another continuity equation.  The probability that,
within a one-dimensional system, there is one electron at position
and spin $X_1\equiv (x_1,\sigma_1)$ and another at $X_2$ is $\left<
\hat{\psi}^{\dagger}(X_1,t)\hat{\psi}^{\dagger}(X_2,t)\hat{\psi}(X_2,t)\hat{\psi}(X_1,t)\right>$.
Here, $\hat{\psi}^{\dagger}(X_1,t) =
e^{iHt/\hbar}\hat{\psi}^{\dagger}(X_1)e^{-iHt/\hbar}$ and
$\hat{\psi}^{\dagger}(X_1)$ creates an electron at $X_1$.  Here and
henceforth, all expectation values are taken with respect to the $N$
particle state of the system $\left|N\right>$.  Averaging over the
position $x_2$ of the second electron, we find from the
Schr\"{o}dinger equation

\bn \lefteqn{\frac{\partial}{\partial t}\int dx_2 \left<
\hat{\psi}^{\dagger}(X_1,t)\hat{\psi}^{\dagger}(X_2,t)
\hat{\psi}(X_2,t)\hat{\psi}(X_1,t)\right>} \n\\
&=&\frac{\partial}{\partial x_1} \left[\frac{\hbar}{2mi}\int dx_2
\left(\frac{\partial}{\partial
x_1} - \frac{\partial}{\partial x_1'}\right) \right.\\
&&\left.\times\left<
\hat{\psi}^{\dagger}(X_1,t)\hat{\psi}^{\dagger}(X_2,t)\hat{\psi}(X_2,t)
\hat{\psi}(X_1',t)\right>\right]_{X_1=X_1'}.\n \en
We are led to the definition of the spin-specific two-particle
current density
\bn \label{Jdef}
&&J(x_1,\sigma_1,\sigma_2,t)\\&=&\frac{e\hbar}{m}\int dx_2{\rm
Im}\left\{\partial_{x_1}\rho_2(X_1,X_2;X_1',X_2,t)\right\}_{X_1=X_1'}\n\en
in terms of the two particle reduced density matrix \bn
\lefteqn{\rho_2(X_1,X_2;X_1',X_2',t) = }\n\\
&&\left<\hat{\psi}^\dagger (X_1,t) \hat{\psi}^\dagger
(X_2,t)\hat{\psi} (X_2',t)\hat{\psi} (X_1',t)\right>. \en
A summation over the two spins would yield a current analogous to
the usual definition of current but for the flow of pairs of
particles rather than single particles; for a single pair of
particles that are in momentum eigenstates we would obtain the
average current density of the pair.

We can expand the the two particle density matrix in terms of energy
eigenstates discussed in Appendix C, \bn
\lefteqn{\rho_2(X_1,X_2;X_1',X_2',t)}\\
&=& \int dE \bar{f}(E) \Psi_E^*(X_1,X_2,t)\Psi_E(X_1',X_2',t) \n \en
where $E$ signifies the available energy of a pair of particles and
$\bar{f}(E)$ is a distribution function for the pair, which we take
to depend only on the total energy $E$ of the pair. As we
demonstrate in Sec. III.C, this is consistent with the fundamental
physical requirement that no current flows in the absence of pumping
or bias.

  In the absence of interaction, the two particle density matrix can
be separated into singlet and triplets spin subspaces. The pair
interactions that we consider do not affect spin by assumption,
hence such a separation would continue to apply. Furthermore, the
energy eigenstates would also be spin eignstates, so that we
factorize the states
$\Psi_E(X_1,X_2,t)=\Psi^{\nu}_E(x_1,x_2,t)\chi_\nu$, where
$\chi_\nu$ is a singlet $\chi_S$ or a triplet
$\chi_{T_{\sigma=0,\pm1}}$ spin state, and effects of the
time-dependent potential is felt only by the spatial part
$\Psi_E(x_1,x_2,t)$.  Thus we write the two particle density matrix
as
\bn
\rho_2=\rho_S\otimes\chi_{_S}\chi_{_S}^{\dagger}+\sum_{\sigma=0,\pm
1} \rho_{T,\sigma} \otimes
\chi_{_{T,\sigma}}\chi_{_{T,\sigma}}^{\dagger}\en 
where $\rho_S=\Psi^*_{E,S}(x_1,x_2,t)\Psi_{E,S}(x_1,x_2,t)$ and
$\rho_{T_\sigma}=\Psi^*_{E,T_\sigma}(x_1,x_2,t)\Psi_{E,T_\sigma}(x_1,x_2,t)$
denote the spatial components.  Energy dependence on the spin states
is implicit. We can then define the current density for each spin
subspace. For singlets we have
\begin{widetext}
\bn\label{current} J_{S}(x,t)&=&\frac{e\hbar}{m}\int dx_2 {\rm
Im}\left\{\partial_{x_1}\rho_S(x_1,x_2;x_1'x_2,t)\right\}_{x_1=x_1'}
\\
&=&\frac{e\hbar}{m}\int dE f(E)\int
dx_2\int\frac{dk_1}{2\pi}\int\frac{dk_2}{2\pi} \ \delta
\left(\frac{\hbar^2k_1^2}{2m}+\frac{\hbar^2k_2^2}{2m}-E\right) {\rm
Im}\left\{\partial_{x_1}
\Psi_{k_1,k_2}(x_1,x_2,t)\Psi_{k_1,k_2}^*(x_1',x_2,t)\right\}_{x_1=x_1'}
\n\en 
\end{widetext}
where we have parameterized the spatially symmetric singlet states
$\Psi_{E,S}(x_1,x_2,t)=\Psi_{k_1,k2}(x_1,x_2,t)$ by single-particle
momenta $k_1$ and $k_2$ corresponding to the noninteracting momentum
eigenstates from which these states can be generated as we show in
the following subsection. The symbol $f(E)$ denotes the state
occupation only within the singlet subspace.

We note that $J_S$ gives the flow of probability of finding one
member of a singlet at $x_1$ irrespective of the location of the
other member. If we average over the position of the other particle,
we will obtain the expectation of the current over the length of the
system.

\subsection{Interacting two particle states}

Suppose our system has the time-independent many-body Hamiltonian,
\bn \label{Ham}
H & = & \int dX \hat{\psi}^\dagger(X)h(x)\hat{\psi}(X)\\
& & + \int dX\int dX'
\hat{\psi}^\dagger(X)\hat{\psi}^\dagger(X')V(x,x')\hat{\psi}(X')\hat{\psi}(X)\n
\en where $h(x) = -\frac{\hbar^2}{2m}\frac{\partial ^2}{\partial
x^2} + W(x)$ is a first quantized single particle Hamiltonian with
some external potential $W(x)$. The equation of motion for
$\Psi_{k_1,k_2}(x_1,x_2,t)$ can be derived directly from the
Schr\"{o}dinger equation.  One finds that with suitable
approximations (see Appendix C) a two-particle equation arises
\bn\label{twobody} [i\hbar \frac{\partial}{\partial t} -
h(x_1)-h(x_2)-V(x_1,x_2)]\Psi_{k_1,k_2}(x_1,x_2,t)=0\n\\
\Rightarrow [E(k_1,k_2)- H_0 - V(x_1,x_2)] \Psi_{k_1,k_2}(x_1,x_2,t)
=0 \h{5mm}\en where
$H_0 = h(x_1)+h(x_2)$ and $\Psi_{k_1,k_2}(x_1,x_2,t)$ has the
trivial time dependence $\Psi_{k_1,k_2}(x_1,x_2,t) =
e^{-iE(k_1,k_2)t/\hbar} \tilde{\Psi}_{k_1,k_2}(x_1,x_2,t)$.

If the one body potential and two body interaction were absent, the solution would
simply take the form of free singlet states
\bn
\Phi_{\bar{k}}(\bar{x})=\frac{1}{\sqrt{2}}[\phi_{k_1}(x_1)\phi_{k_2}(x_2)+
\phi_{k_1}(x_2)\phi_{k_2}(x_1)]\label{free-state}\en 
where $\phi_k$ denotes a single particle momentum eigenstate with
momentum $\hbar k$, and where we have introduced the notation
$\bar{x}\equiv \{x_1,x_2\}$ and $\bar{k}\equiv \{k_1,k_2\}$.  Even
in the presence of a one-body potential, this form still holds
except that the single particle states would be defined by
$h(x)\phi_{k}(x) = E_k\phi_k(x)$.

In the presence of the two body interaction $V(\bar{x})=V(x_1,x_2)$, the
interacting or scattering singlet states $\Psi$ can be expressed in terms of the free state
$\Phi$ by the Lippmann-Schwinger (LS) equation 
\bn\label{LS} \Psi_{\bar k}(\bar{x})&=&\Phi_{\bar k}(\bar{x})+\int d\bar{x}'
G(\bar{x},\bar{x}';E)V(\bar{x}')\Phi_{\bar k}(\bar{x}')\en
where the full retarded two particle Green's function satisfies
\bn
[E-H_0-V(\bar{x})]G(\bar{x},\bar{x}';E)=\delta(\bar{x}-\bar{x}').\n
\en A few comments about expression (\ref{LS}) are in order.
First, due to the two-body interaction, the scattering state
generally involves a range of momenta, so the subscript $\bar k$
there simply serves as a book-keeping label to indicate the
non-interacting state that generates it. Secondly, although the
wavefunctions are singlet functions, the expression involves the
\emph{unsymmetrized} full retarded two particle Green's function
$G(\bar{x},\bar{x}';E)$. Finally, Eq~(\ref{LS}) applies to
\emph{any} two particle wavefunction, regardless of its symmetry.
If we choose a symmetric, spin-independent potential, the symmetry
of the free states is preserved by the scattering and $\Psi_{\bar
k}(\bar{x})$ will have whatever symmetry that $\Phi_{\bar
k}(\bar{x})$ has. In our case, we are interested in the evolution
of a spatially symmetric singlet state, so we apply (\ref{LS}) to
this case.  It is not as interesting to consider the pumping of
triplet wavefunctions since the current could consist of an
unspecified superposition of all three different triplet spin
states.

\section{Pumped Singlet current}

\subsection{Pumping via Interaction}

\begin{figure}[b]
\includegraphics*[width=1.2\columnwidth]{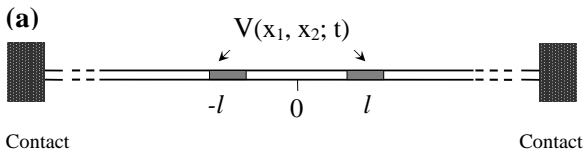}\vspace{-5.3cm}
\caption{(a) Schematic figure of a one dimensional system in which
the \emph{two}-body interaction $V(x_1,x_2;t)$ acts when $x_1$ and
$x_2$ are both within a finite interval near $-l$ or when both are
within a finite interval near $l$. } \label{Fig1}
\end{figure}

Since the equation of motion (\ref{LS}) does not inherently
distinguish between singlet and triplet states, if we wish to pump
only singlets we need to find a potential that only affects
singlets. Obviously, manipulating the external potential $W(x)$ in
(\ref{Ham}) would affect single particle states as well as
singlets.  Thus, we focus on the two-particle interaction
$V(x,x')$.  Within our one-dimensional system we consider a
time-varying two particle interaction that exists at two specific
localized regions, as shown in Fig.~\ref{Fig1}.  Two electrons
interact only when they are both in one such region. If either
electron is outside the regions, there is no interaction, and we
assume that the interaction vanishes between an electron at one
region and an electron at the other region. Physically, if each
region is sufficiently well localized in space, and if the
one-dimensional system consists of a single channel, the Pauli
Exclusion principle will disallow two electrons in a triplet state
from both occupying the same region.  As a result, electrons in a
triplet will not feel the interaction. Only singlets will feel its
effects.  The specifics of how such a localized interaction could
be implemented are discussed in Sec. V. For our derivations below,
we will assume that it is exactly true that the interaction
$V(\bar{x},t)$ at each site affects only singlets, and that
triplets are completely unaffected by it.

\subsection{Adiabatic Perturbation}

The most commonly used description of quantum pumping employs the
well-known Brouwer formula \cite{Brouwer-1} that derives directly
from the Landauer-Buttiker formalism.  Another common description
uses Flouquet theory \cite{Floquet-Buttiker, Floquet-Kim}. Since
those approaches rely upon single particle scattering matrix in 1D,
not appropriate for describing interacting pairs, we have taken an
\emph{a priori} approach rooted in adiabatic perturbation as used by
Thouless \cite{Thouless} in his original paper on quantum pumping.
That approach has been used in some recent papers
\cite{nano-transport} as well. This will merely constitute our point
of departure, because we have the significant complication of
interacting particles with pumping driven by the interaction, while
the previous studies were derived in a single-particle picture.

The singlet current in Eq.~(\ref{current}) is defined in terms of
a time-dependent singlet two particle state $\Psi_{\bar
k}(\bar{x})$. Therefore the determination of the pumped current
reduces to describing the evolution of this singlet
state in the presence of a time-varying interaction $V(\bar{x},t)$ 
\bn H=H_0+V(\bar{x},t)\en
in Eq. (\ref{twobody}).  Here, the free Hamiltonian $H_0$ is still
time-independent and its two-body energy eigenstates are still of the
form shown in Eq.~(\ref{free-state}) comprised of symmetric
combinations of single particle eigenstates.

For an adiabatic process, the potential, or the two-particle
interaction in our case, is assumed to vary slowly compared to the
time spent by the particles in the region of the potential, i.e.
the dwell time is much shorter than the time for variations in the
potential \cite{Buttiker-Landauer}.  As a result, it is
appropriate to insert the ansatz $\Psi_{\bar k}(\bar{x},t) =
e^{-i\int dt E(\bar{k},t)/\hbar}
\tilde{\Psi}_{\bar{k}}(\bar{x},t)$ into Eq. (\ref{twobody}), we
are left with \bn i\hbar \frac{\partial}{\partial t}
\tilde{\Psi}_{\bar k}(\bar{x},t) = [H_0 + V(\bar{x}) -
E(\bar{k},t)]
\tilde{\Psi}_{\bar{k}}(\bar{x},t)\label{adiabatictwobody}.\en
Henceforth we will drop the tilde symbol on $\tilde{\Psi}$; the
phase $e^{-i\int dt E(\bar{k},t)/\hbar}$ that distinguishes $\Psi$
from $\tilde{\Psi}$ does not affect the current (\ref{current}) in
any case.  To find a zeroth order solution, we neglect the time
derivative and solve the instantaneous equation
\bn\label{full-WF-defn}
[E(\bar{k},t)-H_0-V(\bar{x},t)]\Psi_{\bar{k}}^t(\bar{x})&=&0\n\\
\en where the time $t$ is simply a parameter. This equation on
inversion becomes the Lippman-Schwinger equation, Eq~(\ref{LS}), for
the instantaneous scattering state $\Psi_{\bar{k}}^t(\bar{x})$. To
find corrections, we evaluate the instantaneous Green's function
\bn\label{full-GF-defn}
[E-H_0-V(\bar{x},t)]G^t(\bar{x},\bar{x}';E)&=&\delta(\bar{x}-\bar{x}'),
\en and note that the exact solution to (\ref{adiabatictwobody}) is
\bn\label{adiabaticsolution}\Psi_{\bar{k}}(\bar{x},t)=\Psi_{\bar{k}}^t(\bar{x})-i\hbar\int
d\bar{x}' G^t(\bar{x},\bar{x}';E)\frac{\partial}{\partial
t}\Psi_{\bar{k}}(\bar{x}',t).\en Iterating this equation, we can
compute corrections to the zeroth order solution
$\Psi_{\bar{k}}^t(\bar{x})$. To first order,
\bn\Psi_{\bar{k}}(\bar{x},t)&\simeq&\Psi_{\bar{k}}^t(\bar{x})-i\hbar\int
d\bar{x}' G^t(\bar{x},\bar{x}';E)\frac{\partial}{\partial t}\Psi_{\bar{k}}^t(\bar{x}')\n\\
&\equiv&
\Psi_{\bar{k}}^t(\bar{x})+\Delta\Psi_{\bar{k}}(\bar{x},t)\label{wf-lin-ord}\en
Higher orders are left out here because of the assumption of
adiabaticity.  Taking the derivatives of
Eqs~(\ref{full-WF-defn},\ref{full-GF-defn}) with respect to time
yields the following important identities for the time derivatives
of the instantaneous functions
\bn \label{deriv-identities}\dot{\Psi}_{\bar{k}}^t(\bar{x})&=&\int
d\bar{x}'G^t(\bar{x},\bar{x}';E)\dot{V}(\bar{x}',t)\Psi_{\bar{k}}^t(\bar{x}')\\
\dot{G}^t(\bar{x},\bar{x}';E)&=&\int
d\bar{x}''G^t(\bar{x},\bar{x}';E)\dot{V}(\bar{x}'',t)G^t(\bar{x}'',\bar{x}';E)\n\en
which allow us to write the second term in
Eq.(\ref{wf-lin-ord})(the term
first order in the time derivative) as 
\bn\label{1st-order} &&\Delta \Psi_{\bar{k}}(\bar{x},t)=
-i\hbar\int d\bar{x}'
G^t(\bar{x},\bar{x}';E)\dot{\Psi}^t(\bar{x}')
\\
&=&-i\hbar\int d\bar{x}'\int d\bar{x}''
G^t(\bar{x},\bar{x}';E)G^t(\bar{x}',\bar{x}'';E)
\dot{V}(\vec{x}'')\Psi^t(\vec{x}'') \n\en
while the zeroeth order term is simply $\Psi^t_{\bar{k}}(\bar{x})$
determined by the Lippmann-Schwinger equation (\ref{LS}).

\emph{Henceforth, we sometimes simplify notation by writing
$G(\bar{x},\bar{x}')$ instead of $G^t(\bar{x},\bar{x}';E)$.}

\subsection{Zeroth order: No Spontaneous Current}

At the zeroth order in the time-dependence, at each instant the system
is unaware of the fact that the potential is changing.  Since there is
no bias either, the current should vanish, because a non-vanishing
current at this order would essentially be a spontaneous current, which is unphysical. In order to
demonstrate that our definition of the singlet current passes this
crucial test, we explicitly evaluate the current at the zeroth order
in the time-derivative, which is given by
\bn\label{zeroth-1} J_0&=&\frac{e\hbar}{m}\int dE f(E)\int dx_2
\int\frac{dk_1}{2\pi}\int\frac{dk_2}{2\pi}\\&&\times \delta
\left({\textstyle\frac{\hbar^2k_1^2}{2m}+\frac{\hbar^2k_2^2}{2m}}-E\right){\rm
Im}\left\{\Psi_{\bar{k}}^{t*}(\bar{x})
\partial_{x_1} \Psi_{\bar{k}}^t(\bar{x})\right\}\n\en
Using the following identity for the free two particle Green's function
\bn\label{Im-G} {\rm
Im}\{G_0(\bar{x},\bar{x}';E)\}=\h{-1mm}-\pi\h{-2mm}\int\h{-1mm}
d\bar{k} \delta\left({\textstyle\frac{\hbar^2\bar{k}^2}{2m}}-E\right)
\Phi_{\bar{k}}(\bar{x}) \Phi_{\bar{k}}^{*}(\bar{x}')\label{Im-G0}\en
together with the Lippman-Schwinger Eq. (\ref{LS}) we can reduce
the expression~(\ref{zeroth-1}) to

\bn J_0&=&\frac{-e\hbar}{\pi m}\int dE f(E)\int dx_2
{\rm Im}\left[\partial_{x_1}{\rm Im}\{G_0(\bar{x},\bar{x}')\} \right. \n\\
&&+ \int d\bar{x}''
\left\{\partial_{x_1}G(\bar{x};\bar{x}'')\right\}V(\bar{x}''){\rm
Im}\{G_0 (\bar{x}'',\bar{x})\}\n\\ &&+\int d\bar{x}'\left.
G^*(\bar{x};\bar{x}')V(\bar{x}')\partial_{x_1}{\rm Im}\{G_0 (\bar{x},\bar{x}')\}\right.\n\\
&&+\int d\bar{x}'\int d\bar{x}''\left\{ G^*(\bar{x};\bar{x}')V(\bar{x}')\right\} \n\\
&&\h{1cm} \left.\partial_{x_1} \left\{G(\bar{x};\bar{x}'')V(\bar{x}'')
{\rm Im}\{G_0(\bar{x}'',\bar{x}')\right\}\right]\en 
The first term vanishes immediately because it involves the
imaginary part of a real number. Then after some manipulations where
we use the Lippmann-Schwinger equation, the Dyson equation
$G=G_0+G_0VG$, along with the reciprocity property of the Green's
functions, the expression for the current at the zeroeth order can
be reduced to

\bn J_0&=&\frac{e\hbar}{2\pi m}\int dE \ f(E)\int d\bar{x}'\int
d\bar{x}'' V(\bar{x}')V(\bar{x}'')\\&&\times\ {\rm
Im}\left\{iG(\bar{x}'';\bar{x}')\int dx_2\left[G_0
(\bar{x}',\bar{x})\partial_{
x_1}G_0(\bar{x};\bar{x}'')\right]\right.\n\\&&\ \ \ \ -\left.i
G^*(\bar{x}'';\bar{x}')\int
dx_2\left[G_0^*(\bar{x};\bar{x}'')\partial_{
x_1}G_0^*(\bar{x},\bar{x}')\right]\right\}.\n\en
The kernel of the integral is of the form $F\times
I_{GG}(x_1,\bar{x}',\bar{x}'')$ evaluated in
Eq.~(\ref{gg-combination}) in Appendix B (here $F=iG(x',x'';E)$).
Such an integral was shown to vanish for $|x_1|>|x_1'|,|x_1''|$.
This applies here since the current is measured outside the region
of interaction and the variables $x_1'$ and $x_1''$ are associated
with the localized interaction factors $V(\bar{x}')$ and
$V(\bar{x}'')$. Therefore the net current at the zeroeth order
vanishes, $J_0=0$ .

\subsection{First order: Adiabatic Pumped Current}

The pumped current to first adiabatic order is
a bilinear form involving the zeroth order wavefunction $\Psi_{\bar{k}}^t(\bar{x})$ and the first order wavefunction correction $\Delta\Psi_{\bar{k}}(\bar{x},t)$ from Eq. (\ref{wf-lin-ord}):

\bn J_1=\h{-1mm}\frac{e\hbar}{m}\int dE f(E)\h{-1mm}\int\h{-1mm}
dx_2\h{-1mm}
\int\h{-1mm}\frac{dk_1}{2\pi}\h{-1mm}\int\h{-1mm}\frac{dk_2}{2\pi} \
\delta \h{-1mm}
\left({\textstyle\frac{\hbar^2k_1^2}{2m}+\frac{\hbar^2k_2^2}{2m}}-E\right)\n\\\h{-1cm}
\times{\rm Im}\left\{ \Psi_{\bar{k}}^{t*}(\bar{x})\partial_{x_1}
\Delta\Psi_{\bar{k}}(\bar{x},t)+\Delta\Psi_{\bar{k}}^*(\bar{x},t)\partial_{x_1}\Psi_{\bar{k}}^t(\bar{x})
\right\}\h{2mm}\label{1st-order-1}\en

Using the identity in Eq.~(\ref{Im-G0}) and by multiple use of the
Lippmann-Schwinger equation and the Dyson equation, in a
straightforward but lengthy calculation, the above expression can
be  reduced to

\bn J_1&=&-\frac{e\hbar^2}{2\pi m}\int dE f(E)\h{-1mm} \int
d\bar{x}'\int d\bar{x}''\dot{V}(\bar{x}'')\int  dx_2\\
&\times
&\left[\left\{G^*(\bar{x}'',\bar{x})-G(\bar{x}'',\bar{x})\right\}\partial_{x_1}\left\{G(\bar{x},\bar{x}')
G(\bar{x}'\bar{x}'')\right\}
\right.\n\\ &-&\left.\left\{G^*(\bar{x},\bar{x}')
G^*(\bar{x}'\bar{x}'')\right\} \partial_{
x_1}\left\{G^*(\bar{x},\bar{x}'')-G(\bar{x},\bar{x}'')\right\}\right].\n\en
The identity $\int d\bar{x}'
G(\bar{x},\bar{x}')G(\bar{x}',\bar{x}'') =-\partial_E
G(\bar{x},\bar{x}'')$, (Appendix A)  then transforms this to
\begin{widetext}
\bn\label{pumped-current-1} J_1&=&J_{1a}+J_{1b}\n\\ J_{1a}
&=&\frac{\hbar^2}{2\pi m}\int dE \ f(E)\frac{\partial}{\partial E}\left[\int
d\bar{x}''\dot{V}(\bar{x}'') \int dx_2 {\rm
Im}\{G^*(\bar{x}'',\bar{x};E)
\partial_{x_1}G(\bar{x},\bar{x}'';E)\}\right]
\n\\J_{1b}&=&\frac{\hbar^2}{2\pi m}\int dE \ f(E)\int d\bar{x}'\int
d\bar{x}''\dot{V}(\bar{x}''){\rm Im}
\{G^*(\bar{x}',\bar{x}'';E)\int dx_2G^*(\bar{x},\bar{x}';E)\partial_{x_1}G^*(\bar{x},\bar{x}'';E)\n\\
&&\h{6cm}+ G(\bar{x}',\bar{x}'';E)\int
dx_2G(\bar{x}'',\bar{x};E)\partial_{x_1}G(\bar{x},\bar{x}';E)\}\en
\end{widetext}
We will now show that for any observation point $x_1$ greater than
the region of non-vanishing $V$, the term $J_{1b}$ would not
contribute in a periodically varying pumping cycle. We should first
point out that the pair of terms in $J_{1b}$ are \emph{not} complex
conjugates, due to the assymmetry with respect to the exchange of
$\bar{x}'\leftrightarrow \bar{x}''$, so we cannot argue that taking
the imaginary part makes it vanish. Using the Dyson equation we can
make the following expansion
\begin{widetext}
\bn\label{expansion-2nd-term} &&{\rm Im}
\left\{G^*(\bar{x}',\bar{x}'';E){\textstyle {\textstyle \int}}
dx_2G^*(\bar{x},\bar{x}';E)\partial_{x_1}G^*(\bar{x},\bar{x}'';E)+
G(\bar{x}',\bar{x}'';E){\textstyle \int}
dx_2G(\bar{x}'',\bar{x};E)\partial_{x_1}G(\bar{x},\bar{x}';E)\right\}\n\\
=&&{\rm Im}\left\{G^*(\bar{x}',\bar{x}''){\textstyle \int} dx_2
G^*_0(\bar{x},\bar{x}')\partial_{x_1}G^*_0(\bar{x},\bar{x}'')+G(\bar{x}',\bar{x}''){\textstyle
\int}
dx_2G_0(\bar{x},\bar{x}'')\partial_{x_1}G_0(\bar{x},\bar{x}')\right\}\n\\
+&&\h{-3mm} 
{\textstyle \int} d\bar{y}\ {\rm Im}\left\{G^*(\bar{x}',\bar{x}'')
V(\bar{y})G^*(\bar{y},\bar{x}''){\textstyle \int}
dx_2G^*_0(\bar{x},\bar{x}')\partial_{x_1}G^*_0(\bar{x},\bar{y})+G(\bar{x}',\bar{x}'')V(\bar{y})G(\bar{y},\bar{x}''){\textstyle
\int} dx_2G_0(\bar{x},\bar{y})\partial_{x_1}G_0(\bar{x},\bar{x}')\right\}\n\\
+&&\h{-3mm}{\textstyle \int} d\bar{z}\ {\rm
Im}\left\{G^*(\bar{x}',\bar{x}'')G^*(\bar{z},\bar{x}')
V(\bar{z}){\textstyle \int}
dx_2G^*_0(\bar{x},\bar{z})\partial_{x_1}G^*_0(\bar{x},\bar{x}'')+G(\bar{x}',\bar{x}'')V(\bar{z})G(\bar{z},\bar{x}'){\textstyle
\int} dx_2G_0(\bar{x},\bar{x}'')\partial_{x_1}G_0(\bar{x},\bar{z})\right\} \n\\
+&&\h{-3mm}{\textstyle \int} d\bar{z}{\textstyle \int} d\bar{y}\
{\rm Im}\
\left\{G^*(\bar{x}',\bar{x}'')\{V(\bar{y})G^*(\bar{y},\bar{x}')V(\bar{z})G^*(\bar{z},\bar{x}'')\}{\textstyle
\int}
dx_2G^*_0(\bar{x},\bar{y})\partial_{x_1}G^*_0(\bar{x},\bar{z})\right.\n\\&&\h{-3mm}
\left.\h{4cm}+G(\bar{x}',\bar{x}'')\{V(\bar{y})G(\bar{y},\bar{x}'')V(\bar{z})G(\bar{z},\bar{x}')\}{\textstyle
\int} dx_2G_0(\bar{x},\bar{y}) \partial_{x_1}G_0(\bar{x},\bar{z})
\right\}
 \en 
 \end{widetext}
All four terms on the right hand side of the expression above
involve integrals of the form $F\times
I_{GG}(x_1,\bar{x}',\bar{x}'')$ in Eq.~(\ref{gg-combination}) in \
Appendix B, the function $F$ being different in each term. Using the
properties of that integral in the Appendix we see easily that the
last two terms would vanish for $x_1$ outside the interaction
region. In fact if the observation point $|x_1|\rightarrow\infty$,
which would guarantee $|x_1|>|x_1'|$ , by the same argument the
first two terms would be zero as well. However since the variable
$x_1'$ is not associated with any interaction factor $V$, in
principle it is not bounded and we have to allow for $|x_1|<|x_1'|$;
in that case Eq~(\ref{gg-combination-i}) implies that for
observation points outside the interaction region $|x_1|>|x_1''|$ ,
we can reduce the first two terms on the right hand side (which
provide the only non-vanishing contributions) to

\bn
\h{-4mm}&&2\frac{m}{\hbar^2}\{\theta(x_1'-x_1)\theta(x_1-x_1'')-\theta(x_1''-x_1)\theta(x_1-x_1')\}
\n\\\h{-4mm}&&\times{\rm
Im}[G(\bar{x}',\bar{x}'')\{G_0(\bar{x}',\bar{x}'')+{\textstyle \int}
d\bar{y}
G_0(\bar{x}',\bar{y})V(\bar{y})G(\bar{y},\bar{x}'')\}]\n\\
\h{-4mm}&&=2\frac{m}{\hbar^2}{\rm
Im}[G(\bar{x}',\bar{x}'')G(\bar{x}',\bar{x}'')]\n\\
\h{-4mm}&&\times\{\theta(x_1'-x_1)\theta(x_1)-\theta(-x_1)\theta(x_1-x_1')\}\en
Inserting this into the second line of the expression for $J_{1b}$
in Eq~(\ref{pumped-current-1}) and using the identity in
Eq~(\ref{deriv-identities}) we get the following

\bn J_{1b}&=&\frac{1}{\pi}\frac{\partial}{\partial_t}\int
d\bar{x}'\{\theta(x_1'-x_1)\theta(x_1)-\theta(-x_1)\theta(x_1-x_1')\}\n\\&&
\times\int dE
\ f(E){\rm Im} \{G(\vec{x}',\vec{x}';E)\}\\
&=&-\frac{\partial}{\partial_t}\int
d\bar{x}'\rho(\bar{x}')\n\\&&\times
\{\theta(x_1'-x_1)\theta(x_1)-\theta(-x_1)\theta(x_1-x_1')\}\n\en
In the last step we use the definition of the instantaneous
two-dimensional density of states $\frac{1}{\pi}{\rm Im}
\{G(\bar{x}',\bar{x}';E)\}=-\rho(\bar{x}',E)$ and did the integral
over energy to get the local density $\rho(\bar{x}')$. As we see,
this is a total time derivative that would vanish on integrating
over a full period. In any case for an observation point $|x'|\gg 0$
even without doing the time-integral, this term can be made
negligible, and actually in the single particle picture used to
describe charge pumping, the equivalent of this term is identically
zero since the observation point is taken to be asymptotically far
from the region of interaction. Thus we can now write a compact
expression for the net contribution to the adiabatic pumped singlet
current to first order for periodically varying two particle
interaction,
\begin{widetext}
\bn\label{pumped-current-2} J_1(t)&=&\frac{e\hbar^2}{2\pi m}\int dE
\ f(E)\frac{\partial}{\partial E}\left[\int
d\bar{x}'\dot{V}(\bar{x}') \int dx_2 {\rm
Im}\{G^*(\bar{x}',\bar{x};E)
\partial_{x_1}G(\bar{x},\bar{x}';E)\}\right].\en
\end{widetext}

\section{Validity for the non-interacting case:
Charge pumping}

Our result has the advantage that it also describes adiabatic
quantum pumping where the independent particle description is
assumed.  We simply interpret the functions as single-particle
objects, i.e. the green's functions are single particle green
functions, $G(\bar{x}',\bar{x}'';E)\rightarrow g(x',x'';E)$, and we
then do not have the integral over $x_2$, so that
\bn &&J(t) =\frac{e\hbar^2}{2\pi m}\int dE \
f(E)\frac{\partial}{\partial E}\\&&\h{1cm}\times\left[\int
dx'\dot{V}(x') {\rm Im}\{g^*(x',x;E)
\partial_{x}g(x,x';E)\}\right]
\n\en 
Noting that the single particle Green's function has the
asymptotic form
\bn g(x,x'';E)=\frac{m}{ik\hbar^2}e^{ikx}\psi^*(x')\en 
where $k=\sqrt{2mE/\hbar^2}$, one obtains
\bn &&J(x,t) =\frac{em}{2\pi \hbar^2 }\int dE \
f(E)\frac{\partial}{\partial E}\left\{\frac{1}{k}\langle\psi|\dot{V}|\psi\rangle\right\}\en 
which agrees with the expression in
Ref.[\onlinecite{nano-transport}] on noting that in that paper the
states are normalized by $\sqrt{k}$.  That expression in turn has
been shown to be equivalent to the Brouwer formula
\cite{Brouwer-1}.

\begin{figure}[b]
\includegraphics*[width=1.2\columnwidth]{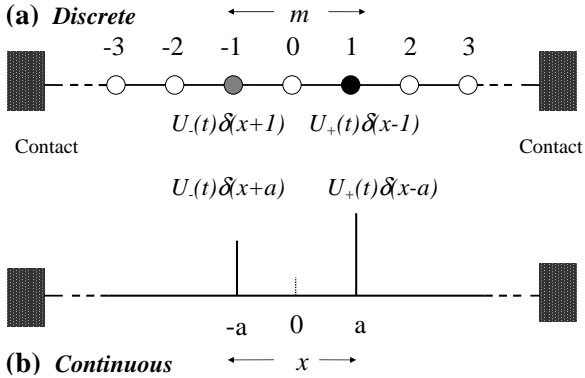}\vspace{-2.5cm}
\caption{(a) A discrete model for a singlet pump comprising of a
chain of quantum dots where  the interaction between a singlet pair
is non-vanishing only when both particles are together at one of the
two shaded dots. (b) The continuum version of the same model where
the pair interaction is non-zero only at the two points $x=\pm a$
indicated by spikes. The strengths of the two-particle interaction
at the two sites are $U_{\pm}(t)$ which are the time-varying
parameters. At any instant the two parameters can be of different
magnitude indicated by different shades in the discrete model and
different heights of the spikes in the continuum model (the short
dotted line is simply a marker for the origin $x=0$). } \label{Fig2}
\end{figure}

\section{Physical model for a Singlet Pump: A lattice of quantum dots}

In theoretical studies of quantum pumping, a turnstile model
\cite{Levinson-turnstile} has been used, wherein delta function
potentials exist at two points in a 1D system.  The cyclic variation
of the strength of the potentials serves as the pumping cycle.  The
region between the two spikes can be interpreted as a scattering
region, such as a quantum dot, and the strength of the potentials a
measure of the barriers segregating the scattering region from the
rest of the system (e.g. the leads). We can adapt the turnstile
model, proposing a singlet pump in which delta function external
potentials acting at the two points are replaced with the
\emph{electron-electron interaction} acting only at those points. In
a continuum of positions, this model of extremely localized
interactions seems somewhat unrealistic.  However, in a 1D
tight-binding lattice, we would have interactions at two lattice
sites, which seems physically reasonable and appropriate for the
application of our results. Physically such a lattice model could be
implemented, for instance, by a chain of quantum dots where precise
voltage controls produce only time-varying interaction between the
members of an electron pair when both particles occupy one of two
specific dots. A schematic of the model is shown in Fig.~\ref{Fig2}.

Formally the expressions in the discrete case are identical with
those of the continuum case, as we have verified explicitly
\cite{Das-PRL}: The Green's functions in the continuum case need
only be reinterpreted as discrete Green's functions and spatial
integrals need only be replaced by spatial sums.  Since the general
derivation in this paper has been in a continuum model, we will
derive the results for the turnstile model also in a continuum form.

The interaction is 
\bn V(\bar{x},t)=U_-(t)\delta(\bar{x}+\bar{a})+
U_+(t)\delta(\bar{x}-\bar{a})\label{hubbard-pot}\en 
where $\pm\bar{a}\equiv \pm \{a,a\}$.  The strength of the delta
functions $U_-$ and $U_+$ are the two time-dependent pumping
parameters. The instantaneous pumped singlet current in Eq~(\ref
{pumped-current-2}) then becomes 
\bn \label{current-delta-1} J_S(t) &=&\frac{e\hbar^2}{2\pi m}\int dE
f(E)\sum_{\bar{y}=\pm \bar{a}}\dot{U}_{\pm}(t)
\n\\&&\times\partial_E{\rm Im}\left[\int dx_2G^*(\bar{x},\bar{y})
\partial_{x_1}G(\bar{x},\bar{y})\right]\en 
The full instantaneous Green's function for this system can be
written compactly as
\bn\label{gf-expansion} G(\bar{x},\bar{y})=
G_0(\bar{x},\bar{y})+F_-(\bar{y})
G_0(\bar{x},-\bar{a})+F_+(\bar{y})G_0(\bar{x},\bar{a}) \n\en 
with the coefficients being 
\bn F_\pm(\bar{y})=\left[\frac{T_\pm G_0(\pm\bar{a},\bar{y})+T_\pm
T_\mp G_0(2\bar{a})G_0(\mp\bar{a},\bar{y})}{1-T_\mp T_\pm
G_0(2\bar{a})G_0(2\bar{a})}\right].\en 
We have introduced the T-matrix for a single Hubbard
interaction (a single term in Eq.~(\ref{hubbard-pot})),
\bn T_\pm(t)=[U_\pm^{-1}(t)+G_0(0)]^{-1}.\en
We have also introduced a shorthand for the free Green's functions:
Since the free Green's functions depend only on the absolute
difference of the coordinates,
$G_0(\bar{x},\bar{y})=G_0(|\bar{x}-\bar{y}|)$, we define
$ G_0(2\bar{a})=G_0(\bar{a},-\bar{a})=G_0(-\bar{a},\bar{a})$ and
$G_0(0)=G_0(\bar{a},\bar{a})=G_0(-\bar{a},-\bar{a})$. 
Using the expansion Eq.~(\ref{gf-expansion}) we can evaluate the
coordinate integrals in Eq~(\ref{current-delta-1}) to be

\bn &&\frac{\hbar^2}{m}\int dx_2G^*(\bar{x},\pm\bar{a})
\frac{\partial}{\partial_{x_1}}G(\bar{x},\pm\bar{a})=\\
\h{-7mm}&&\h{-7mm}-{\rm Im}\{G_0(0)\}\left[1+[F_\pm(\vec{a})+F_\pm^*(\vec{a})]+|F_\mp(\vec{a})|^2+|F_\mp(\vec{a})|^2\right]\n\\
\h{-7mm}&&\h{-7mm}-{\rm Im}\{F_\mp(\vec{a}) G_{0R}(2a) -
F_\mp^*(\vec{a})
G_{0R}^*(2a)\}\n\\
\h{-7mm}&&\h{-7mm}-{\rm
Im}\{F_\pm^*(\vec{a})F_\mp(\vec{a})G_{0R}(2a)-F_\mp^*(\vec{a})F_\pm(\vec{a})G_{0R}^*(2a)
\}\n\en
In order to obtain this form  we used the expansion
Eq~(\ref{gf-expansion}) above to reduce the above integral to a sum
of integrals of the form shown in Eqs.~(\ref{int-g*g-1})and
(\ref{g*g-combination}) considered in the Appendix B. Then we used
their respected evaluated forms given in
Eqs.~(\ref{g*g-combination-g}) and (\ref{g*g-combination-0}). What
is noteworthy is that the expressions depend on $G_{0R}(2a)$, the
regular part of the two-particle Green's function which has no
singularity.  This is discussed in the Appendix in Eq.~(\ref{2Dgf})
After carrying through a lengthy but straightforward set of
algebraic manipulations, of the above expression, we insert the
result into  Eq~(\ref{current-delta-1}) to get the following
expression for the singlets pumped in a single cycle of period
$\tau$
\begin{widetext} \bn  Q_S(\tau)=\frac{-e}{2\pi}\int_0^\tau\h{-2mm}dt
\h{-1mm}\int \h{-1mm}dE f(E)\frac{\partial}{\partial E}\sum_{\pm}
\dot{U}_{\pm}(t) \frac{|T_\pm(t)|^2\left[ {\rm
Im}\{G_0(0)\}(1+|T_\mp(t)G_0(2\bar{a})|^2)\pm 2{\rm
Im}\{T_\mp(t)G_0(2\bar{a})G_{0R}^{\pm}(2\bar{a})\}
\right]}{U_\pm(t)^{2}|1-T_\mp(t)T_\pm(t)G_0(2\bar{a})G_0(2\bar{a})|^2}.\en
\end{widetext}
where $G_{0R}^{+}(2\bar{a})=G_{0R}(2\bar{a})$ and
$G_{0R}^{-}(2\bar{a})=G_{0R}^*(2\bar{a})$. Thus in order to evaluate
the current we only need to evaluate the \emph{free two-particle
lattice} Green's function, for only two specific arguments
$G_0(0)=G_0(\bar{a},\bar{a};E)$ and
$G_0(2\bar{a})=G_0(\bar{a},-\bar{a};E)$. Exact analytical forms
exist for the lattice Green's functions, $G_0$, in terms of elliptic
integrals \cite{Economou}.

As mentioned above the expression for the discrete lattice is
identical to this with the replacement of the Green's functions by
their discrete equivalents $G_0(0)\rightarrow
G_0(\bar{m},\bar{m};E)$ and $G_0(2\bar{a})\rightarrow
G_0(\bar{m},-\bar{m};E)$ where $\pm m$ correspond to the lattice
sites where the time varying interaction occurs.  In
Fig.~\ref{Fig2}(a), we take $m=1$.  The effect of changing the value
of $m$ on the number singlet pairs pumped per cycle was presented in
an earlier paper \cite{Das-PRL}.

\section{Characteristics of a Quantum dot for
time-varying interaction}

While the physical model comprising of a chain of quantum dots as
presented above yields a measurable singlet current in theory, the
actual experimental implementation requires a feasibility study
involving realistic quantum dots. We will now suggest a specific
configuration for an experimentally realizable quantum dot to tailor
it for singlet pumping. In the rest of the paper, we will establish
that the characteristics of our proposed configuration will allow
the variation and manipulations to satisfy the requirements of a
singlet pump.

\begin{figure}[t]\vspace{-1cm}
\includegraphics*[width=1.1\columnwidth]{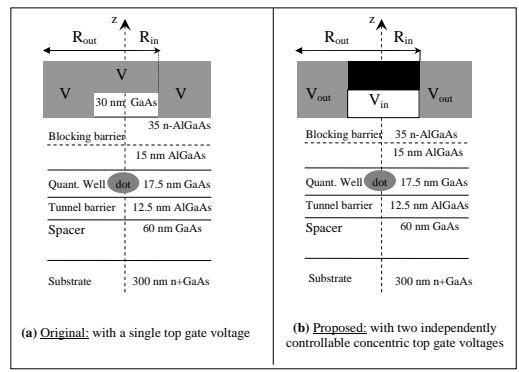}\vspace{-6cm}
\caption{(a) Schematic figure of the quantum dot used in Refs.
[\onlinecite{Ashoori}] and [\onlinecite{Bednarek}]. Here $V$ is the
potential energy at the top gate. (b) Our proposed modification of
that structure with two concentric metal gates, with independently
controllable voltages; $V_{in}$ is the net potential \emph{energy}
of an electron at the top edge of the blocking barrier ($V_{in}=-e
\times \Phi_{in}$, where $\Phi_{in}$ is the voltage) right below the
inner gate (darker shade).  Likewise $V_{out}$ is the potential
energy of an electron at the base of the outer gate (lighter
shade).} \label{Fig3}
\end{figure}

To maintain close contact with experiments, we specifically consider
a cylindrically symmetric GaAs/Al$_x$Ga$_{1-x}$As quantum dot used
in experiments reported in Ref.~\onlinecite{Ashoori}.  To adapt
the dot for singlet pumping, we propose a single modification: instead
of being one single disk, the top gate needs to be made of an inner
disk of radius $R_{in}$ and a concentric outer annulus with inner
and outer radii $R_{in}$ and $R_{out}$, so that the voltages
$V_{in}$ and $V_{out}$ on the two pieces can be controlled
independently.  This provides two independent parameters for
manipulating the shape and depth of the lateral potential.  A
schematic cut-away view is shown in Fig.~\ref{Fig3}, in the original
experiment (a) and with our proposed modification (b).

Along the direction of the z-axis (Fig.~\ref{Fig3}),
electron confinement is provided by the difference in conduction
band edge energies between GaAs and Al$_x$Ga$_{1-x}$As.  The result is a
quantum well of width 17.5 nm in the middle GaAs layer. Lateral
confinement within that GaAs layer is due to the inhomogeneous
electric potential generated by the top gate.  In the original
setup, a 30 nm thick GaAs cylindrical cap over the center of the dot
created that inhomogeneity, while in our modified version, the
different voltages on the inner and outer discs both play a role in the lateral confinement.
The inhomogeneity is such that the potential energy of an
electron V ($=-e\times \Phi$ where $\Phi$ is the gate voltage) is
lower near the axis of the dot so that electrons are attracted
to the region under the inner disc or cap.

We determine the potential profile within the dot by numerically
solving the Poisson equation
\bn (\nabla_R^2+\partial_z^2) V= \left\{\begin{array}{cr}0&0<z<105\ {\rm nm}\\
(-e)\times-\frac{n_D e}{K\epsilon_0}&105<z<140\ {\rm nm}\end{array}\right.\en
applying the methods and parameters used in Ref.~\onlinecite{Bednarek}
to match experimental observed capacitance spectroscopy peaks of
Ref.~\onlinecite{Ashoori}.  Here, $z$ is measured from the top of the
substrate, so that the region $105 \ {\rm nm} < z < 140 \ {\rm nm}$ is the charged
blocking barrier. In the Poisson equation, $K=12.85$ is the dielectric
constant for undoped GaAs, and the charge density in the blocking
barrier is taken to be $n_D=4.62\times 10^{17}$ cm$^{-3}$. We take the
Fermi energy of the bottom electrode as the energy reference, and
specify $V_{in}$ and $V_{out}$ at the top surface of the
cylinder. Thus, at $z=140$ nm, $V_{in}$ is taken to be a positive
constant potential energy in the inner disk of radius $R_{in}$ and
$V_{out}$ is a positive constant potential energy on the outer disc
between $R_{in}$ and $R_{out}$ (to create electron confinement we must
have $V_{in}<V_{out}$). The potentials are defined to be those at the
edge of the blocking barrier, $z=140$ nm, so that $V_{in}$ and
$V_{out}$ include the Schottky potential of about $0.65$ eV at the
metal gate semiconductor interface, as well as the drop due to the GaAs
cap under the inner disk. The boundary condition on the curved outer
surface of the cylindrical dot is set by assuming a large value of $R_{out}$, so that the
electric field is essentially parallel to the curved surface and
aligned along the z-direction. We solve the 1D Poisson equation in the
z-direction to determine the potential profile on this outer surface
at $r=R_{out}$
\bn V(R_{out},z)= \left\{\begin{array}{cr} B_1 z+B_2& z<105\ {\rm nm}\\
+\frac{n_D e^2}{K\epsilon_0}\frac{z^2}{2} +A_1 z +A_2& z>105\
{\rm nm}\end{array}\right..\en
Applying the boundary conditions at the top ($V=V_{out}$) and bottom
surface ($V=0$) and matching solutions at the bottom edge of the
blocking barrier layer at $z=105$ nm determines the constants 
$ A_1=(V_{out}-9.9684)/140\ {\rm eV/nm}, B_1=(V_{out}-0.3987)/140\
{\rm eV/nm}, A_2=3.5886\ {\rm eV}, {\rm and}\ B_2 = 0$.

\begin{figure}[t]\vspace{-4cm}
\includegraphics*[width=\columnwidth]{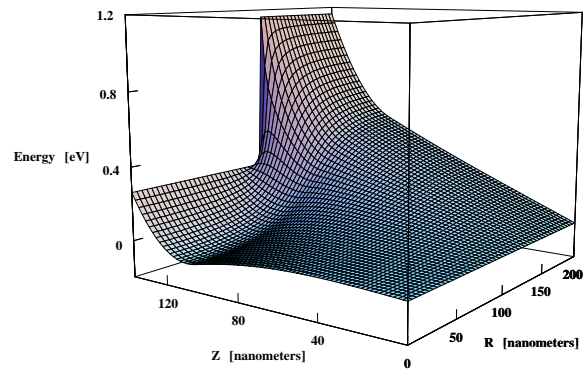}\vspace{-1cm}
\caption{Potential energy of an electron in the structure depicted
in Fig.~\ref{Fig3}(b), without the 0.22 eV offsets due to the AlGaAs
layers. } \label{Fig4}
\end{figure}

\begin{figure}[t]\vspace{-5cm}
\includegraphics*[width=\columnwidth]{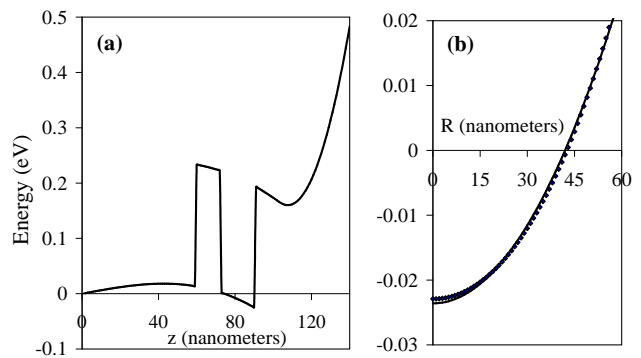}\vspace{-.5cm}
\caption{Radial and axial profiles of the potential along lines
passing through the center of the quantum dot obtained from the
solution of the Poisson equation for the structure depicted in
Fig.~\ref{Fig3}(b). (a) Axial (z) profile along the cylinder axes
(R=0) including the 0.22 eV band offset (b) Radial profile at z=90
nm where the potential is the lowest; the black dots correspond to
the solution of the Poisson equation, and the continuous line is the
parabolic fit.} \label{Fig5}
\end{figure}

With the value of the potential on the boundary determined, the full
2D Poisson equation in $r,z$ is solved using FISHPACK FORTRAN
routines for the solution of separable elliptic partial differential
equations \cite{FISHPACK}. Figure~\ref{Fig4} plots the shape
of the potential  \emph{without} the 0.22 eV conduction band offset of the
AlGaAs; we took $R_{in}=120$ nm and $R_{out}=400$ nm.
Fig.~\ref{Fig5}(a) shows the computed z-profile along the cylinder
axis at $R=0$ nm now \emph{with} the 0.22 eV AlGaAs offset
included; this shows that the region where
electrons are trapped resembles a finite square well along the z-axis.
Fig.~\ref{Fig5}(b) shows the computed profile, in the
middle of the finite square well ($z=90$ nm), in the radial direction.
The solid line is a parabolic fit, which shows
clearly that in the radial direction the trapped electrons feel a
harmonic oscillator potential.

\section{An Electron-pair in the modified quantum dot}

The problem of two electrons in a quantum dot has been analyzed in
detail in the literature.  As we see from Fig.~\ref{Fig5}, the
potential in our quantum dot is well approximated by a finite square
well in the z-direction and a harmonic oscillator in the radial
direction. The effective Hamiltonian operator for two electrons is
then
\bn \hat{H}=-\frac{\hbar^2}{2m^*}\left[\nabla^2_{r1}+\partial_{z1}^2
+\nabla^2_{r2}+\partial_{z2}^2\right] +V_{sq.
well}\n\\+\frac{e^2/4\pi\epsilon_0
K}{(r_1-r_2)^2+(z_1-z_2)^2}+\frac{1}{2}m^*\omega^2( r_1^2+r_2^2)
 \label{SE} \en
In the numerical estimates that follow, we neglect the
$z-$dependance in the Coulomb interaction in order to separate the
radial and axial directions in the Schrodinger equation.
Quantitatively, the neglect of this term in the denominator of the
Coulomb interaction will cause a slight overestimate of the Coulomb
interaction energy, an effect that can be diminished if the dot is
significantly narrower along the z-axis than in the radial
direction. However, qualitatively this approximation makes no
difference for the demonstration of quantum pumping feasibility.

The total energy of an electron in the dot can now be separated into
two parts $E=E_z+E_R$, an axial quantum well energy $E_Z$ which is
well approximated by a finite square well and a transverse energy
$E_R$ that can be well approximated using a harmonic oscillator.

\subsection{Axial energy, $E_z$}

The size of the quantum well is $L=17.5$ nm. and the height is given
by $U=|U_0|+0.22$ eV, where $0.22$eV is the band offset between GaAs
and Al$_x$Ga$_{1-x}$As and $U_0$ is the depth of the well below the
Fermi energy of the bottom lead, our reference energy. For an
infinite well of the same width we get:
$ E_z^{\infty}=\frac{2\hbar^2}{m^*L^2}
\times\left(\frac{\pi}{2}\right)^2=18.31 {\rm meV}$
taking $m^* = 0.067m$ for undoped GaAs. For a finite square well,
the ground state energy is determined by
\bn E_z=\frac{2\hbar^2x^2}{m^*L^2}  = (7.43 {\rm meV}) \times
x^2;\h{.2cm} x\tan(x)&=&\sqrt{P^2-x^2}\n\en with $P^2=134.8\times U
{\rm (in\ eV)}$ where $U$ is the well depth. We found that as $U$
varies over the range (0.22, 0.26) \ eV,  $E_z$ remains in the range
(13.0, 13.4)\rm meV. Since it turns out that $U_0$ changes by less
than $0.01$eV when we vary the parameters $V_{out}$ and $V_{in}$,
the value of $E_Z$, which is defined relative to the bottom of the
well, changes by less than $\sim 1\%$.  Therefore, in our
calculations, the variation of $E_Z$ relative to the bottom of the
well can be neglected. As Fig.~\ref{Fig5}(a) shows, the bottom of
the well is not completely flat, varying from $0.22$eV to
$|U_o|+0.22$eV.  However, for the same reason, this slope has little
effect on the constancy of $E_Z$.

\subsection{Radial energy, $E_R$}

The transverse or radial Hamiltonian can be separated into
center-of-mass (R) and relative coordinates (r) \cite{Zhu}
\bn H_{com}&=&-\frac{1}{2}\nabla^2_R+\frac{1}{2}\gamma^2R^2\n\\
H_{rel}&=&-2\nabla^2_r+\frac{1}{8}\gamma^2r^2+ \frac{2}{r}\en
where energies are scaled in Rydbergs $R_D=\frac{\hbar^2}{2m^*
a_D^2}$ and the length scale is
$a_D=\hbar^2(4\pi\epsilon_0K)/(m^*e^2)$ and $\gamma=2(a_D/l_0)^2$
where $l_0 = \sqrt{\hbar/m^*\omega}$ is the oscillator length.  In
these units, the harmonic oscillator energy spacing, $\hbar\omega$,
is given by $R_D\gamma$. \emph{Excluding} the Coulomb interaction
the solutions are those for a 2-dimensional harmonic oscillator with
circular symmetry with energy
\bn E_{R}-E_{int}=(2N+1+|M|)\gamma+ (2n+1+|m|)\gamma.\en
The parity of the spatial wavefunction  is determined by the
relative coordinate quantum number $m$.  For $m=0$, the spatial
wavefunction is even under particle change ($r \rightarrow -r$), so
the state must have an antisymmetric spin wavefunction, a singlet.
For $m=1$ the spatial wavefunction is odd under particle
interchange, so the state must have a symmetric spin wavefunction, a
triplet.  We conclude that the lowest energy states are the singlet
with $(N,M,n,m)=(0,0,0,0)$ and the triplet with
$(N,M,n,m)=(0,0,0,1)$. These have energies $2\hbar\omega$ and
$3\hbar\omega$ respectively \emph{without} including the Coulomb
interaction energy $E_{int}$.

The interaction energy $E_{int}$ has been calculated for 2D circular
quantum wells in the literature \cite{Zhu,McKinney} for values of
$\gamma \sim 0.1 - 1$. We use those results for $E_{int}$ to
determine the total radial energy $E_{R}$.

\section{Experimental Feasibility of Pumping Entangled Electrons}

\subsection{Desired scenario for singlet pumping}

We would like to vary the top gate potential energies $V_{in}$ and
$V_{out}$ in such a way that the shape of the lateral confining
potential changes with the following constraints:

(i) At all times, the singlet state is energetically accessible, but
the triplet state is too high in energy to be occupied.

(ii) The energy of a single electron in the dot with respect to the
Fermi energy $E_F=0$ of the bottom electrode stays constant and
\emph{far off resonance} with $E_F$.  This suppresses single
electron tunneling, which depends strongly on including the resonant
region in the pumping cycle \cite{Levinson-turnstile}.

(iii) There is significant variation of the Coulomb interaction between the two electrons: the singlet energy can be varied appreciably to permit pumping of pairs.\\

We present numerically computed results for two sample
configurations of the dot corresponding to different specific values
of the top gate potential energies $V_{in}$ and $V_{out}$ which
satisfy the above criteria. These two configurations could serve as
points within a cyclic variation of the top gate potential energies.

The lowest energy state spin singlet (with quantum numbers
$(N,M,n,m)=(0,0,0,0)$) and spin triplet (with quantum numbers
$(N,M,n,m)=(0,0,0,1)$) have total energies:
\bn 1s:\h{.5cm} E^{(S)}(\omega)=2U_0+2E_z+2\hbar\omega+E_{int}^{(S)}(\omega)\n\\
2p:\h{.5cm}
E^{(T)}(\omega)=2U_0+2E_z+3\hbar\omega+E_{int}^{(T)}(\omega)
\label{singlettriplet}\en
In order to satisfy condition (i) we look for values of $\omega$
such that the total triplet energy lies above the Fermi level i.e.
the reference energy in our calculations, while the $E^{(S)}$ lies
below:
\bn E^{(S)}(\omega)<0\h{1cm}  E^{(T)}(\omega)>0\en
In order to satisfy condition (ii) to maintain the single particle
energy levels fixed with respect to the Fermi level, we have to
simultaneously ensure that in any two configurations (1 and 2), we
have:
\bn\hbar\omega_1+U_0^{(1)}+E_z=\hbar\omega_2+U_0^{(2)}+E_z \label{singleelectronenergy},\\
\Rightarrow\ \hbar\omega_1-\hbar\omega_2=U_0^{(2)}-U_0^{(1)}.\n\en

\begin{widetext}\hspace{2cm}
\begin{table}[h]
\caption{Summary of numerically computed parameters for two specific
configurations of the potential profile of the proposed quantum dot,
that would satisfy the requirements of singlet pumping. The two
configurations would be part of a continuous and cyclic variation of
the potential profile of the dot.}
\begin{tabular}{lrrr}
&&$Configuration\ 1$ \h{6mm}\vline& $Configuration\ 2$ \\ \hline
&\vline & \vline &\\
 &$V_{out} $\h{1cm}\vline &$1158.1\ meV$\h{6mm} \vline& $950.6\ meV$ \\
 &$V_{in} $\h{1cm}\vline &  $263.1\ meV$\h{6mm} \vline& $301.2\ meV$\\\hline
 &$l_0=\sqrt{\hbar/m^*\omega} $\h{1cm}\vline &  $14.35\ nm$\h{6mm} \vline&
 $15.55
\ nm$\\
 &$\hbar\omega $\h{1cm}\vline &  $5.5\ meV$\h{6mm} \vline& $4.7
 meV$\\
 &$U_0 $\h{1cm}\vline &  $-22.9\ meV$\h{6mm} \vline& $-22.1\ meV$ \\
 &$E_Z $\h{1cm}\vline &  $13.2\ meV$\h{6mm} \vline& $13.2\ meV$ \\
 Single Electron level& $U_0+E_Z+\hbar\omega $\h{1cm}\vline &
 $-4.2\ meV$\h{6mm} \vline& $-4.2\ meV$ \\
 \hline
 Singlet interaction energy&$E_{int}^{(S)} $\h{1cm}\vline &$7.3\ meV$\h{6mm} \vline& $6.6\ meV$ \\
 Triplet interaction energy&$E_{int}^{(T)} $\h{1cm}\vline &$4.6\ meV$\h{6mm} \vline& $4.2 meV$ \\
 &$E_{int}^{(T)}-E_{int}^{(S)} $\h{1cm}\vline &$-2.7\ meV$\h{6mm} \vline& $-2.4\ meV$\\
 \hline
 Radial Singlet energy&$E_{R}^{(S)} $\h{1cm}\vline &$18.4\ meV$\h{6mm} \vline& $16.0 meV$ \\
Radial Triplet energy& $E_{R}^{(T)} $\h{1cm}\vline &$21.2\ meV$\h{6mm} \vline& $18.3\ meV$ \\
 &$E_{R}^{(T)}-E_{R}^{(S)} $\h{1cm}\vline &$2.8\ meV$\h{6mm} \vline& $2.3\ meV$\\
 \hline
 Total Singlet energy \ \ &$2U_0+2E_z+E_R^{(S)} $\h{1cm}\vline &$-0.5\ meV$\h{6mm}
  \vline& $-0.9\ meV$ \\
 Total Triplet energy \ \ &$2U_0+2E_z+E_R^{(T)}$\h{1cm}
 \vline &$+0.9\ meV$\h{6mm} \vline& $+0.25\ meV$ \\
\end{tabular}
\end{table}
\end{widetext}

\begin{figure}[b ]\vspace{-2cm}
\includegraphics*[width=1.1\columnwidth]{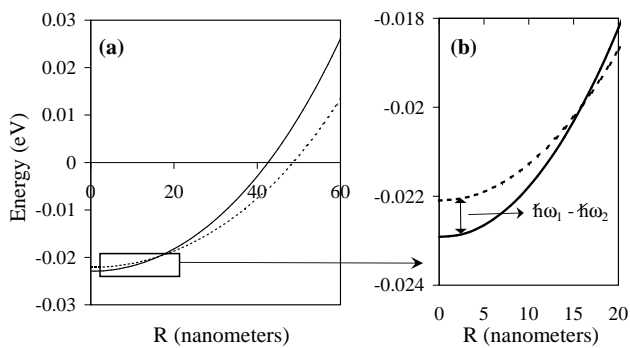}\vspace{-5cm}
\caption{(a) Radial profiles through the lowest potential region of
the dot for two distinct set of top gate potential energies:
continuous line $\rightarrow (V_{out},V_{in})=(1158.1\ {\rm meV},
263.1\ {\rm meV})$ and dotted line
$\rightarrow(V_{out},V_{in})=(950.6\ {\rm meV}, 301.2\ {\rm meV})$.
They correspond to the two configurations described in Table I.  (b)
Expanded view of the lowest energy region. The shift $U_0$ between
the two configurations is $\hbar\omega_1-\hbar\omega_2$, keeping the
single particle ground state energy unchanged.} \label{Fig6}
\end{figure}

\subsection{Experimental parameters for singlet pumping}

We take the dimensions of our quantum dot to be similar to those of
Ref. \onlinecite{Ashoori}, except that our dot has a slightly smaller
radius of $120$ nm for the inner top gate while estimates
\cite{Bednarek} found that in the experiment the radius of the GaAs
cap was about $200$ nm.

We explicitly identify two sample locations in parameter space that
satisfy all of our conditions.  (To find a complete pumping path
through parameter space is straightforward since one has two
parameters $V_{out}$ and $V_{in}$ and only one precise quantitative
constraint (ii) above).  Say the two sample locations have radial
oscillator energies $\hbar\omega=5.5 $ meV and $\hbar\omega=4.7 $ eV
(corresponding to $\gamma=1$ and $\gamma=0.85$ respectively).  Then
the well depths must compensate such that the single electron energy
(\ref{singleelectronenergy}) remains constant.  For example, the
locations could satisfy $U_0=-22.9 $ meV and $U_0=-22.1 $ meV so
that the single electron energy is held at $-4.2 $ meV, constant and
strongly off resonance with the Fermi energy. At these values of
$\hbar \omega$ and $U_0$, the quantum dot geometry is such that the
interaction energies \cite{Zhu} for singlets are $7.3 $ meV and $6.6
$ meV respectively and for triplets $4.6 $ meV and $4.2 $ meV,
respectively. Then equation (\ref{singlettriplet}) implies that the
total energies for the singlets are $-0.5 $ meV and $-0.9 $ meV for
the two locations in parameter space while the total energies of the
triplets are $0.9 $ meV and $0.25 $ meV respectively. Thus, the
singlet states are energetically accessible while the triplet states
are not.

Our simulation shows that the two configurations are realized in the
dot at $(V_{out}, V_{in})=(1158.1 {\rm meV}, 263.1 {\rm meV})$ and
$(V_{out}, V_{in})=(950.6 {\rm meV}, 301.2 {\rm meV})$, from our
numerical computation, which can be rounded off to the appropriate
significant figures.  Figure~\ref{Fig6}(a) shows the radial profile
in the dot for the two configurations. Configuration 1 has a higher
difference between $V_{in}$ and $V_{out}$, leading to a larger value
of $\omega$ and causing tighter lateral confinement. Configuration 2
has a lower difference between $V_{in}$ and $V_{out}$ leading to
smaller value of $\omega$. However, because $V_{in}$ is higher than
in configuration 1, the well is not as deep. This can be clearly
seen in Fig.~\ref{Fig6}(b) where the profiles for the two
configurations cross due to the compensating effects of increased
$\omega$ and lowered $U_0$.

Table I summarizes the results of our simulation. Energetically the
singlets are accessible in both configurations, while
triplets are inaccessible (point (i) above).  The energy of the
single electron level in the dot is fixed at $-4.2 {\rm meV}$
significantly off-resonant with $E_F=0$ (point (ii) above).  The
variation of the Coulomb interaction is about $11\%$ between the two
singlet configurations (point (iii) above).

\section{Conclusions}

In this paper we address several issues related to developing a
quantum singlet pump. We will conclude by providing a summary of our
main results. We have considered the problem of applying the
mechanism of adiabatic quantum pumping to generate a flow of singlet
pairs of electrons, while suppressing the flow of triplets and
uncorrelated single particles. The first challenge was to find the
appropriate theoretical description.  We first identified an
appropriate definition of the current for singlet pairs in analogy
with the current of a stream of uncorrelated electrons, by using the
reduced two-particle density matrix.  We then showed how in the
presence of two-body interactions, the evolution of the many
electron system can be effectively described by the evolution of a
two-particle state particularly when the interaction is spatially
localized.  We confirmed that our definition of current gives a
non-zero current in the absence of bias or time variation.  We then
wrote an adiabatic perturbation expansion for the time varying two
particle states, where the rate of change of the interaction
$\partial V/\partial t$ is assumed slow compared to the time scale
of the dwell time of the particles in the interaction region.  Using
the assumption of adiabaticity, we computed the pumped current to
first order in the time derivative (or equivalently the frequency of
the time-varying interaction). The interaction is assumed to affect
only singlets and is non-vanishing only in a finite region, but
otherwise completely general. By using several identities and
relations for two particle Green's function, that we present in the
appendices, we reduced the expression for the current to a compact
and relatively simple form involving only the instantaneous two
particle Green's functions. The current is seen to have an
additional transient term which we show vanishes identically for a
complete cycle, the term being a total time derivative.

Having established a general but simple expression for the pumped
singlet current due to the action of a time-varying local two body
interaction, we apply it to a specific model.  We consider a
lattice, where the interaction acts and varies in time only at two
of the sites;  the interaction acts only when both particles are
localized on the same site, while the two sites are sufficiently
separated that interaction among an electron at one site with one at
the other can be neglected.  If the sites are well localized in
space, the effect of the interaction on triplets is automatically
suppressed due to vanishing or at least substantial suppression of
the triplet spatial wavefunction at each of the two sites due to the
Pauli principle.   We computed the singlet current for the model as
an exact analytical expression, which has been confirmed to have the
same form for discrete and continuous models.

Finally we show that this model can be implemented using a chain of
quantum dots.  We take the design specifications of a quantum dot
that had been fabricated and studied in the lab and propose one
minor modification that would introduce two concentric top gates
with independently controllable voltages, something that can be
achieved without much difficulty with the methods of fabrication
available currently.  We computed the potential profile of such a
dot with available experimental parameters and computed the energy
of two electrons in such a dot. We showed that by varying the two
gate voltages, (i) significant singlet current can be generated (ii)
triplets can be made energetically inaccessible for states in the
dot, and therefore affected much less by the variation of the
interaction (iii) the single particle states can be maintained at
the same energy and far off-resonance, thereby suppressing current
of uncorrelated single electrons.  Thus in effect such a dot can be
used in a chain to implement our model for generation of a
measurable singlet current.

\acknowledgments

We gratefully acknowledge the support of the Research Corporation,
and support from a Faculty Research Grant from Fordham University.
It is also a pleasure to acknowledge valuable discussions with Ari
Mizel and Sungjun Kim, and helpful advice on the numerical
simulations from G. Recine.

\appendix

\section{Greens functions: Representations and Identities}

The Green's function for the potential-free time-independent
Schrodinger equation is defined by
\bn[E+\textstyle{\frac{\hbar^2}{2m}}\nabla^2]G_0(\bar{x}',\bar{x}'';E)
=\delta(\bar{x}'-\bar{x}'')\en
where $\bar{x}\equiv\{x_1,x_2,\cdots,x_n\}$ denotes the spatial
degrees of freedom, the total energy $E$ shared among them is
assumed real, and the subscript in $G_0$ indicates the absence of
any potential. The degrees of freedom have equivalent
interpretations as $n$ spatial dimensions or as coordinates of $n$
individual particles in one dimension. We study interacting
particles in 1D, but we use `n-particle' or `$n$D' interchangeably
in referring to the Green's functions. The retarded and advanced
Green's functions will be denoted using superscripts $G^+$ and
$G^-$. When the Green's functions obey
$G_0^+(\bar{x}',\bar{x}'';E)=[G_0^-(\bar{x}',\bar{x}'';E)]^*$, we
leave out the superscripts, $G^+\equiv G$ and  $G^-\equiv G^*$.

Our interest being in two body interactions we only need single
particle (or 1D) and two particle (or 2D) Green's functions. The
uppercase $G(\bar{x}',\bar{x}'';E)$ will be reserved for 2D
Green's function with $\bar{x}\equiv \{x_1,x_2\}$, and 1D Green's
functions will be distinguished by using lowercase $g_0(x,x';E)$.
Single-particle eigenstates are likewise denoted by lowercase
$\phi_k(x)$; expansion in such eigenstates readily establishes the
following useful identities:
\bn\label{gf-identities} \int_{-\infty}^{\infty} dx
\phi_k^*(x)g_0(x,x';E)
&=&\frac{\phi_k^*(x')}{E-E_k+i\eta}
\\
\int_{-\infty}^{\infty} d\epsilon
\frac{g_0(x',x'';E-\epsilon)}{\epsilon-E_p+i\eta} &=&
-2i\pi g_0(x',x'';E-E_p)\n\\
\int_{-\infty}^{\infty} d\epsilon
\frac{g_0(x',x'';E-\epsilon)}{\epsilon-E_p-i\eta} &=& 0\n\en
\bn\int_{-\infty}^{\infty} dx
g_0^*(x,x';E_1)g_0(x,x'';E_2)=\h{2.8cm}
\n\\\frac{g_0^*(x'',x';E_1)
-g_0(x',x'';E_2)}{E_2-E_1+i\eta} \n\\
\int_{-\infty}^{\infty} dx g_0(x,x';E_1)g_0(x,x'';E_2)=\h{2.9cm}
\n\\\frac{g_0(x'',x';E_1)
-g_0(x',x'';E_2)}{E_2-E_1}\stackrel{E_2\rightarrow
E_1}{\rightarrow} -\frac{\partial}{\partial_{E_1}}
g(x',x'';E_1)\n\en

For scattering problems the appropriate eigenstates are plane
waves which in 1D are $\phi_k(x)=e^{ikx}$; they provide a
coordinate representation for the free 1D Green's function for
real energies $E$
\bn\label{1Dgf}
&&g_0^{\pm}(x',x'';E)=\int_{-\infty}^{\infty}\frac{dk}{2\pi}\frac{e^{ik(x'-x'')}}
{E-(\hbar^2k^2/2m)\pm i\eta}
\\&=&\mp\frac{i\theta(E)e^{\pm i\sqrt{\frac{2m}{\hbar^2}E}|x'-x''|}}{\frac{\hbar^2}{m}\sqrt{2mE/\hbar^2}}
-\frac{\theta(-E)e^{-\sqrt{\frac{2m}{\hbar^2}|E|}|x'-x''|}}{\frac{\hbar^2}{m}\sqrt{2m|E|/\hbar^2}}
\n\en 
The difference of the retarded and the advanced Green's functions
gives the two-point correlation function
\bn\label{1D-dos} \rho(x',x'';E)&=&\frac{1}{-2\pi i}
\left[g_0^+(x',x'';E)-g_0^-(x'',x';E)\right] \\&=&
\theta(E)\int_{-\infty}^{\infty}\frac{dk}{2\pi}
\delta(E-(\hbar^2k^2/2m))e^{ik(x'-x'')}\n\en 
Performing the momentum integral demonstrates consistency with the
definition of $g_0^\pm$. In the case of equal coordinate arguments a
different contour integral is involved, but the end result agrees
with simply setting $x'=x''$ in the above expressions;
$\rho(x',x';E)$ defines the one-dimensional density of states.

The 2D free Green's function can be written in terms of the 1D
Green's functions
\bn\label{2D-1D} G_0^\pm(\bar{x}'\bar{x}'';E)&&=\\\h{-3mm}&&\h{-3mm}
\frac{\pm i}{ 2\pi}\int_{-\infty}^{\infty} d\epsilon
g_0^\pm(x_1',x_1'';E-\epsilon)g_0^\pm(x_2',x_2'';\epsilon) \n\en 
Using an eigenfunction expansion in Cartesian coordinates and
integrating out one momentum component gives
\bn\label{2Dgf} &&G_0^\pm(\bar{x}'\bar{x}'';E) =\\
 &&\mp i\int_{0}^{\sqrt{\frac{2mE}{\hbar^2}}} \frac{dk}{2\pi}
\frac{2\cos[k(x_2'-x_2'')]e^{\pm
i\sqrt{2mE/\hbar^2-k^2}|x_1'-x_1''|}}{\frac{\hbar^2}{m}\sqrt{2mE/\hbar^2-k^2}}
\n\\&&-\int_{\sqrt{\frac{2mE}{\hbar^2}}}^{\infty}
\frac{dk}{2\pi}\frac{2\cos[k(x_2'-x_2'')]
e^{-\sqrt{k^2-2mE/\hbar^2}|x_1'-x_1''|}}{\frac{\hbar^2}{m}\sqrt{k^2-2mE/\hbar^2}}.\n\en
The first term has both real and imaginary parts and is always
regular ($R$) as a function of the coordinate arguments, while
second term is always real and is singular ($S$) when
$\bar{x}'=\bar{x}''$; so we name the two terms 
$ G_{0R}^\pm(\bar{x}'\bar{x}'';E)$ and
$G_{0S}^\pm(\bar{x}'\bar{x}'';E)$ respectively.

\section{Integrals used in computing current}

We will apply the results of the preceding appendix to evaluate
the generic expressions involving two particle or 2D Green's
functions required in computing current. The 2D Green's functions
below are at energy $E$, not explicitly shown for the sake of
compact notation.

\subsection{Integral with conjugate Green's functions}

We encounter expressions involving a pair of conjugate 2D Green's
functions when computing the current:

\bn\label{int-g*g-1}I_{G^*G}(\bar{x}',\bar{x}'')=
\int_{-\infty}^{\infty}
dx_2G_0^*(\bar{x},\bar{x}')\frac{\partial}{\partial
x_1}G_0(\bar{x},\bar{x}'')\en 
By expressing the 2D Green's functions above in terms of 1D
Green's functions as in Eq.~(\ref{2D-1D}), and then using the
identities in Eq.~(\ref{gf-identities}) we can reduce it to the
form

\bn\label{int-g*g-2}
I_{G^*G}(x_1,\bar{x}',\bar{x}'')&=&\frac{i}{2\pi}\int_{-\infty}^{\infty}
d\epsilon [g(x_2',x_2'';\epsilon)-g^*(x_2',x_2'';\epsilon)]\n\\
\h{-1cm}&&\h{-1cm}\times
g^*(x_1,x_1';E-\epsilon)\frac{\partial}{\partial
x_1}g(x_1,x_1'';E-\epsilon).\en

Inserting the expressions for 1D Green's functions from
Eqs.~(\ref{1Dgf}) and (\ref{1D-dos}) leads to an explicit functional
form that is piecewise continuous, in which the free variable $x_1$
determines the boundaries of continuity; thus for the
\emph{exterior} region $|x_1|>|x_1'|,|x_1''|$, which is relevant to
us, the expression is
\bn\label{int-g*g-3}
\h{-.5cm}&&I_{G^*G}(x_1,\bar{x}',\bar{x}'')=\\\h{-.5cm}&&\pm\left[i\int_{0}^{\sqrt{\frac{2mE}{\hbar^2}}}\frac{dk}{2\pi}
\frac{2\cos[k(x_2'-x_2'')]e^{\pm
i\sqrt{\frac{2m}{\hbar^2}E-k^2}(x_1'-x_1'')}}{\frac{\hbar^4}{m^2}\sqrt{2mE/\hbar^2-k^2}} 
\right.\n\\\h{-.5cm}&&\left.-\int_{\sqrt{\frac{2mE}{\hbar^2}}}^{\infty}\frac{dk}{2\pi}
\frac{2\cos[k(x_2'-x_2'')]e^{-\sqrt{k^2-\frac{2m}{\hbar^2}E}|2x_1-(x_1'+x_1'')|}}{\frac{\hbar^4}{m^2}\sqrt{k^2-2mE/\hbar^2}}\right]\n\en
where the $+$ sign applies for $x_1>x_1',x_1''$ and the $-$ sign for
$x_1<x_1',x_1''$.  We can exploit the similarity of this expression
with $ G_{0}^\pm(\bar{x}'\bar{x}'')$ in Eq.~(\ref{2Dgf}) because we
always encounter it in the functional combination 
\bn\label{g*g-combination} F\times I_{G^*G}(x_1,\bar{x}',\bar{x}'') =\h{4cm}\\
{\rm Im}\left\{F\int_{-\infty}^{\infty}
dx_2G_0^*(\bar{x},\bar{x}')\frac{\partial}{\partial
x_1}G_0(\bar{x},\bar{x}'')\right.\n\\\left.+F^*\int_{-\infty}^{\infty}
dx_2G_0^*(\bar{x},\bar{x}'')\frac{\partial}{\partial
x_1}G_0(\bar{x},\bar{x}')\right\} \n\en 
where $F$ is a complex-valued function and $\bar{x}'\leftrightarrow
\bar{x}''$ are exchanged between the two terms;  the second term in
Eqs.~(\ref{int-g*g-3}), being (i) manifestly real and (ii)
unaffected by the exchange $\bar{x}'\leftrightarrow \bar{x}''$,
would not contribute to Eq.~(\ref{g*g-combination}).  Therefore,
\emph{in this particular combination} the integrals can be replaced
by $ G_{0R}^\pm(\bar{x}'\bar{x}'')$. We need to note that the
exponential in Eq.~(\ref{2Dgf}) contains the \emph{absolute}
difference of the coordinate arguments while
 Eq.~(\ref{int-g*g-3}) does not; this in effect determines the choice
of  $G_{0R}^\pm$ depending on whether $x'>x''$ or $x'<x''$: 
\bn\label{g*g-combination-g} {\rm For}\
x_1>x_1',x_1'',\h{3mm}F\times I_{G^*G}(\bar{x}',\bar{x}'')=\h{2.3cm}\\
\frac{m}{\hbar^2}{\rm Im}\left\{
-[FG_{0R}(\bar{x}',\bar{x}'')-F^*G_{0R}^*(\bar{x}',\bar{x}'')]\theta(x_1'-x_1'')
\right.\n\\\left.\h{4mm} 
+[FG_{0R}^*(\bar{x}',\bar{x}'')-F^*G_{0R}(\bar{x}',\bar{x}'')]\theta(x_1''-x_1')]\right\}
\n\\
{\rm For}\ x_1<x_1',x_1'',\h{3mm}F\times
I_{G^*G}(\bar{x}',\bar{x}'')=\h{2.3cm}\n\\\frac{m}{\hbar^2}{\rm
Im}\left\{
-[FG_{0R}^*(\bar{x}',\bar{x}'')-F^*G_{0R}(\bar{x}',\bar{x}'')
]\theta(x_1'-x_1'')]\right.\n\\\left. \h{4mm}
+[FG_{0R}(\bar{x}',\bar{x}'')
-F^*G_{0R}^*(\bar{x}',\bar{x}'')]\theta(x_1''-x_1')]\right\} \n\en

It is also useful to note that the imaginary part satisfies  
\bn\label{imag-part} {\rm
Im}\left\{I_{G^*G}(\bar{x}',\bar{x}'')\right\}&=&\left\{\begin{array}{lr}
-\frac{m}{\hbar^2}{\rm Im}\{G_{0}(\bar{x}',\bar{x}'')\}& x_1>x_1',x_1''\\
+\frac{m}{\hbar^2}{\rm Im}\{G_{0}(\bar{x}',\bar{x}'')\} &
x_1<x_1',x_1''\end{array}\right.\n\en 
which helps simplify the case $\bar{x}'=\bar{x}''$, when the first
terms in both Eqs.~(\ref{int-g*g-3}) and (\ref{2Dgf}) become
imaginary
\bn\label{g*g-combination-0} F\times
I_{G^*G}(x_1,\bar{x}',\bar{x}')\h{4cm}\n\\=\left\{\begin{array}{lr}-
\frac{m}{\hbar^2}{\rm Im}\{G_0(\bar{x}',\bar{x}')\}{\rm
Re}(F)&x_1>x_1'
\\
+\frac{m}{\hbar^2}{\rm Im}\{G_0(\bar{x}',\bar{x}')\}{\rm Re}(F)&
x_1<x_1'
\end{array}\right..\en

\subsection{Integral with like Green's functions}

We also encounter expressions with a pair of similar 2D Green's
functions
\bn\label{int-gg-1}I_{GG}(x_1,\bar{x}',\bar{x}'') =
\int_{-\infty}^{\infty}
dx_2G_0(\bar{x},\bar{x}')\frac{\partial}{\partial
x_1}G_0(\bar{x},\bar{x}'')\en 
which, by using  Eq.~(\ref{2D-1D}) and the identities in
Eq.~(\ref{gf-identities}), can be reduced to the form
\bn\label{int-gg-2} =\frac{i}{2\pi}\int_{-\infty}^{\infty}
d\epsilon g(x_2',x_2'';\epsilon)
g(x_1,x_1';E-\epsilon)\n\\\times\frac{\partial}{\partial
x_1}g(x_1,x_1'';E-\epsilon)\en
Then the expressions for the 1D Green's functions in
Eq.~(\ref{1Dgf}) provide explicit functional forms which, as in the
previous case, depends on the value of $x_1$. For the
\emph{exterior} region $|x_1|>|x_1'|,|x_1''|$ this is
\bn\label{int-gg-e} I_{GG}(x_1,\bar{x}',\bar{x}'')&=& \pm
\frac{i}{2\pi}\int_{-\infty}^{\infty} \!\!d\epsilon
g_\epsilon(x_2',x_2'')\\&& \left[
\frac{e^{i\sqrt{2m(E-\epsilon)/\hbar^2}|2x_1-(x_1'+x_1'')|}}{i\frac{\hbar^4}{m^2}\sqrt{2m(E-\epsilon)/\hbar^2}}\theta(E-\epsilon)
\right.\n\\&&\left.-\frac{e^{-\sqrt{2m(\epsilon-E)/\hbar^2}|2x_1-(x_1'+x_1'')|}}
{\frac{\hbar^4}{m^2}\sqrt{2m(\epsilon-E)/\hbar^2}}\theta(\epsilon-E)\right]\n\en 
where the $+$ sign applies for $x_1>x_1',x_1''$ and the $-$ sign for
$x_1<x_1',x_1''$. This integral is also of interest for
\emph{interior} region when $|x_1|$ is between $|x_1'|$ and
$|x_1''|$:

\bn\label{int-gg-i}
I_{GG}(x_1,\bar{x}',\bar{x}'')=\left\{\begin{array}{lr}
+\frac{m}{\hbar^2}G_0(\bar{x}',\bar{x}'')&x_1'>x_1>x_1''\\
-\frac{m}{\hbar^2}G_0(\bar{x}',\bar{x}'')&x_1''>x_1>x_1'
\end{array}\right.\en

The following combination is relevant
\bn\label{gg-combination} F\times I_{GG}(x_1,\bar{x}',\bar{x}'') =\h{3cm}\\
{\rm Im}\left\{F\int_{-\infty}^{\infty}
dx_2G_0(\bar{x},\bar{x}')\frac{\partial}{\partial
x_1}G_0(\bar{x},\bar{x}'')\right.\n\\
\left.+F^*\int_{-\infty}^{\infty}
dx_2G_0^*(\bar{x},\bar{x}'')\frac{\partial}{\partial
x_1}G_0^*(\bar{x},\bar{x}')\right\}. \n\en 
For the exterior region $I_{GG}(x_1,\bar{x}',\bar{x}'')$ is
invariant under exchange of coordinate arguments, so the
expression above vanishes, $F\times
I_{GG}(x_1,\bar{x}',\bar{x}'')=0 $, for the \emph{exterior} region
$|x_1|>|x_1'|,|x_1''|$.  In the \emph{interior} region
$I_{GG}(x_1,\bar{x}',\bar{x}'')$ simply changes sign under such an
exchange, so that for 
\bn\label{gg-combination-i} F\times I_{GG}(\bar{x}',\bar{x}'')
=\pm\frac{m}{\hbar^2} {\rm
Im}\left\{FG_0(\bar{x}',\bar{x}'')-c.c.\right\}
\n\\
=\left\{\begin{array}{lr}+ 2\frac{m}{\hbar^2}{\rm
Im}\left\{FG_0(\bar{x}',\bar{x}'')\right\},&x_1'>x_1>x_1''\n\\
-2\frac{m}{\hbar^2}{\rm
Im}\left\{FG_0(\bar{x}',\bar{x}'')\right\},&x_1''>x_1>x_1'
\end{array}\right.\n\\\en

\section{An Effective Two Particle State for the Many body system}

As soon as we consider pair interaction, in principle, we have to
allow for interaction between all possible pairs in the system. In
this Appendix we reduce a full many body system with pair
interaction to an effective description in terms of a pair of
interacting particles both moving in the background of the
interaction field due to all the other electrons in the system.
This allows us to describe the pumping of singlets in terms of the
evolution of a singlet wavefunction very much like the pumping of
individual electrons by a single particle state.

Using the composite notation for position and introduced in Sec.
II, $X=(x,\sigma)$, the Heisenberg equation of motion of a Fermion
creation operator with the Hamiltonian (\ref{Ham}) is
\bn
i\hbar\partial_t\hat{\psi}^\dagger(X,t)=h(x)\hat{\psi}^\dagger(X,t)\h{3cm}\n\\+\int
dX'\hat{\psi}^\dagger(X,t)\hat{\psi}^\dagger(X',t)V(x,x')\hat{\psi}(X',t).\label{dtpsi}\en
Consider $\langle
E_0,N|\hat{\psi}^\dagger(X_1,t)\hat{\psi}^\dagger(X_2,t)|E_2,N-2\rangle$,
a matrix element where $|E_0,N\rangle$ is the N particle ground
state of the system, and $|E_2,N-2\rangle $ is an energy eigenstate
with $N-2$ particles.  Using Eq.~(\ref{dtpsi}), we find that the
equation of motion of this matrix element is
\begin{widetext}
\bn \h{-2.5cm}&&\h{-2.5cm}i\hbar\partial_t\langle
E_0,N|\hat{\psi}^\dagger(X_1,t)\hat{\psi}^\dagger(X_2,t)|
E_2,N-2\rangle =(h(x_1)+h(x_2))\langle
E_0,N|\hat{\psi}^\dagger(X_1,t)\hat{\psi}^\dagger(X_2,t)|E_2,N-2\rangle
\n\\\h{2cm}&&\h{2cm}+\int dX' V(x_1,x')\langle
E_0,N|\hat{\psi}^\dagger(X_1,t)\hat{\psi}^\dagger(X',t)\hat{\psi}(X',t)\hat{\psi}^\dagger(X_2,t)|E_2,N-2\rangle
\n\\\h{2cm}&&\h{2cm} +\int dX' V(x_2,x')\langle
E_0,N|\hat{\psi}^\dagger(X_1,t)\hat{\psi}^\dagger(X_2,t)\hat{\psi}^\dagger(X',t)\hat{\psi}(X',t)|E_2,N-2\rangle
\label{2particleSchrodinger}\en
\end{widetext}
Since both $|E_0,N\rangle$ and $|E_2,N-2\rangle $ are energy
eigenstates, we can replace $i\hbar\partial_t\rightarrow E=E_2-E_0$
on the left hand side of the above equations. Furthermore,
\bn&&\hat{\psi}^\dagger(X',t)\hat{\psi}(X',t)\hat{\psi}^\dagger(X_2,t)\\
=&&\delta(X'-X_2)\hat{\psi}^\dagger(X',t)+\hat{\psi}^\dagger(X_2,t)\hat{\psi}^\dagger(X',t)\hat{\psi}(X',t)
\n\en
due to the Fermion commutation relations; here we denoted
$\delta(X'-X_2) \equiv \delta_{\sigma',\sigma_2}\delta(r'-r_2)$.
Making a mean-field approximation, with the density of background
electrons at $X'$ defined by, $n(X',t) = \langle
E_0,N-2|\hat{\psi}^\dagger(X',t)\hat{\psi}(X',t)|E_0,N-2\rangle$, we
define the two particle state
\bn \Psi_E(X_1,X_2,t)\!=\!\langle
E_0,N|\hat{\psi}^\dagger\!(X_1,t)\hat{\psi}^\dagger\!(X_2,t)|
E_2,N\!-\!2\rangle\ \  \en
and bring (\ref{2particleSchrodinger}) into the form

\bn &&E\Psi_E(X_1,X_2,t)\\
&=& [h(x_1)+h(x_2)+V(x_1,x_2)]\Psi_E(X_1,X_2,t) \n \\
&& +\int dX' [V(x_1,x')+V(x_2,x')] n(X',t)\Psi_E(X_1,X_2,t) \n\en

The last line gives the mean-field influence of the rest of the
electrons with the particular pair of electrons at $X_1$ and $X_2$.
We can include this in the one body potential that each particle
experiences: $ h(x) + \int dX' V(x,x') n(X',t)\rightarrow h(x)$.
Since the two-body interaction $V$ does not affect spin, we can
factorize out the spin part of the wave function. Considering
specifically the singlet subspace and  by introducing the
parametrization in terms of the single particle momentum labels
$k_1, k_2$ we are led to  Eq~(\ref{twobody}).

\end{document}